\documentclass[a4paper,12pt]{article}
\usepackage{jcappub}
\usepackage[symbol]{footmisc}
\usepackage{amsmath,amssymb}
\usepackage{revsymb}
\usepackage{orcidlink}
\usepackage{bm}
\newcommand{\nc}{\newcommand*}

\nc{\Eq}[1]{Eq.~\eqref{#1}}     
\nc{\Fig}[1]{Fig.~\ref{#1}}     
\nc{\Table}[1]{Table~\ref{#1}}  
\nc{\Sec}[1]{Sec.~\ref{#1}}     
\def\({\left(}
\def\){\right)}
\def\[{\left[}
\def\]{\right]}
\def\e{\begin{equation}}
\def\q{\end{equation}}
\def\m{\begin{eqnarray}}
\def\n{\end{eqnarray}}

\begin{document}

\title{Search for a Gravitational-Wave Background from Sound Speed Resonance from Advanced LIGO and Advanced Virgo's First Three Observing Runs}

\author[a]{You~Wu,\orcidlink{0000-0002-9610-2284}}
\author[b,c,*]{Zu-Cheng Chen,\note{Corresponding author.}\orcidlink{0000-0001-7016-9934}}
\author[d,*]{Lang~Liu\orcidlink{0000-0002-0297-9633}}

\affiliation[a]{College of Mathematics and Physics, Hunan University of Arts and Science, Changde, 415000, China}
\affiliation[b]{Department of Physics and Synergetic Innovation Center for Quantum Effects and Applications, Hunan Normal University, Changsha, Hunan 410081, China}
\affiliation[c]{Institute of Interdisciplinary Studies, Hunan Normal University, Changsha, Hunan 410081, China}
\affiliation[d]{Faculty of Arts and Sciences, Beijing Normal University, Zhuhai 519087, China}

\emailAdd{youwuphy@gmail.com}
\emailAdd{zuchengchen@hunnu.edu.cn}
\emailAdd{liulang@bnu.edu.cn}

\abstract{
We search for a stochastic gravitational-wave background (SGWB) originating from scalar-induced gravitational waves (SIGWs) with the sound speed resonance (SSR) effect using data from Advanced LIGO and Advanced Virgo's first three observing runs. The SSR mechanism, characterized by an oscillating sound speed squared term, can induce a nonperturbative parametric amplification of specific perturbation modes during inflation, leading to enhanced primordial curvature perturbations and a significant SIGW signal. We perform a Bayesian analysis to constrain the model parameters describing the SGWB spectrum from the SSR effect. Our results show no statistically significant evidence for the presence of such a signal in the current data. Consequently, we place an upper limit of $|\tau_0| \lesssim 5.9 \times 10^3\,\mathrm{s}$ at $95\%$ confidence level on the start time of the oscillation in the SSR model. These results demonstrate the capability of current gravitational wave detectors to probe inflation models through the SSR mechanism and paves the way for future searches with improved sensitivity.
}
	
\maketitle

\section{Introduction}

The detection of gravitational waves (GWs) by Advanced LIGO~\cite{LIGOScientific:2014pky} and Advanced Virgo~\cite{VIRGO:2014yos} has opened a new window into the universe and ushered in the era of GW astronomy~\cite{LIGOScientific:2018mvr,LIGOScientific:2020ibl,LIGOScientific:2021djp}. These groundbreaking observations have provided unprecedented insights into the dynamics of compact binary systems, such as binary black hole and neutron star mergers~\cite{LIGOScientific:2016aoc,LIGOScientific:2017vwq}. The successful detection of individual GW events has been a remarkable achievement; nevertheless, the observation of another important GW source, the stochastic GW background (SGWB), remains an ongoing quest.

The SGWB is a superposition of unresolved GW signals from various astrophysical and cosmological sources~\cite{Christensen:2018iqi}. Astrophysical sources~\cite{Regimbau:2011rp} contributing to the SGWB include compact binary coalescences~\cite{Zhu:2011bd,Zhu:2012xw} and core-collapse supernovae~\cite{Crocker:2017agi}. On the cosmological side, primordial density perturbations during inflation~\cite{Starobinsky:1979ty,Turner:1996ck,Bar-Kana:1994nri}, cosmic phase transitions~\cite{Kibble:1980mv,Witten:1984rs,Mazumdar:2018dfl}, cosmic strings~\cite{Kibble:1976sj,Sarangi:2002yt,Damour:2004kw,Siemens:2006yp,LIGOScientific:2021nrg}, and cosmic domain walls~\cite{Vilenkin:1982ks,Sikivie:1982qv} in the early universe are potential contributors to the SGWB~\cite{Caprini:2018mtu}. The detection of the SGWB would provide valuable information about the early universe and the properties of high-energy physics beyond the Standard Model.

Among the various sources of the SGWB, scalar-induced GWs (SIGWs), accompanying the formation of primordial black holes (PBHs), have emerged as a promising candidate. PBHs are hypothetical black holes that could have formed in the early universe from the collapse of large density perturbations~\cite{Carr:1974nx,Carr:1975qj}. The formation of PBHs is often associated with the production of SIGWs, as the large curvature perturbations generated during inflation can induce GWs at second order in perturbation theory~\cite{Ananda:2006af,Baumann:2007zm}.

Recently, a novel mechanism for PBH formation and the production of SIGWs, known as sound speed resonance (SSR), has been proposed~\cite{Cai:2018tuh,Cai:2019jah,Chen:2019zza,Chen:2020uhe,Zhou:2020kkf,Jin:2023wri}. 
The SSR mechanism is characterized by an oscillating sound speed squared term, which induces a nonperturbative parametric amplification of specific perturbation modes during inflation. As a result, the power spectrum of primordial curvature perturbations features distinctive peaks at small scales, while preserving near-scale invariance on larger scales, in agreement with the predictions of inflationary cosmology. Moreover, according to second-order cosmological perturbation theory, the amplified primordial curvature perturbations generated by the SSR mechanism are anticipated to give rise to significant GW signals.

Data from LIGO and Virgo's first three observing runs have been used to search for and constrain various potential sources of the SGWB \cite{KAGRA:2021mth,Jiang:2022uxp, Yu:2022xdw,Jiang:2023qht,Jiang:2024aju,Guo:2023gfc,Yuan:2022bem,Jiang:2022svq,Chen:2024mwg,Jiang:2022mzt}, including ultralight scalar and tensor dark matter~\cite{Guo:2023gfc,Yuan:2022bem}, cosmic domain walls~\cite{Jiang:2022svq}, and inflation with null energy condition violation~\cite{Chen:2024mwg}. These analyses demonstrate the power of the LIGO-Virgo network to probe fundamental physics and cosmology through GWs.

In this paper, we explore the constraints on SSR models using GW data from the first three observing runs of Advanced LIGO and Advanced Virgo. We concentrate on a specific scenario proposed in~\cite{Cai:2018tuh}, which incorporates an SSR phase in the early Universe. This model predicts a substantial enhancement of the power spectrum of curvature perturbations, resulting in a significant GW signal that can potentially be probed by current and future GW observations. The remainder of this paper is organized as follows. In Section~\ref{SSR}, we begin by reviewing the SGWB arising from the SSR mechanism. Section~\ref{data} outlines the methodology employed to derive constraints on the model parameters using data from Advanced LIGO and Advanced Virgo. Finally, in Section~\ref{conclusion}, we summarize our findings and discuss the implications of our results.

\section{\label{SSR}SGWB from SSR}
In this paper, we will briefly review the mechanisms underlying SSR. To begin, we introduce the variable $u$, defined as $u \equiv z \zeta$, where $z \equiv \sqrt{2 \epsilon} a / c_s$. Here, $a$ denotes the scale factor, $c_s$ represents the sound speed, and $\epsilon$ is the slow-roll parameter defined as $\epsilon \equiv - \dot H / H^2$. Additionally, $\zeta$ represents the comoving curvature perturbation, which remains invariant under gauge transformations. Considering the perturbation of Einstein equations, the Fourier mode of the variable $u$ satisfies the following equation~\cite{Armendariz-Picon:1999hyi,Garriga:1999vw}:
\begin{equation}\label{eq:uk}
    u''_{k} + \left( c_s^2 k^{2} - \frac{z''}{z} \right) u_{k} = 0,
\end{equation}
where the prime denotes a derivative with respect to the conformal time $\tau$. In the SSR model, the oscillating sound speed is defined as follows~\cite{Cai:2018tuh}:
\begin{equation}\label{eq:cs}
    c_s^2 = \left\{
    \begin{aligned}
        &1, &\tau < \tau_0,\\
        &1 -2 \xi \left[ 1 - \cos \left( 2 k_* \tau \right) \right], &\tau > \tau_0,
    \end{aligned}
    \right.
\end{equation}
where $\xi$ represents the amplitude of the oscillation, $k_*$ corresponds to the oscillation frequency, and $\tau_0$ is the start time of the oscillation. The oscillating sound speed can be realized within the effective field theory of inflation~\cite{Achucarro:2010da, Achucarro:2012sm}.
To ensure that $c_s^2 \geq 0$, we must impose the condition $\xi < 1/4$. Furthermore, $k_*\tau_0$ is chosen to be a multiple of $\pi$ to guarantee a smooth transition of the sound speed from its constant value to the oscillating regime. Consequently, the effective mass during the oscillation period can be simplified as
\begin{equation}
    \frac{z''}{z} = \frac{2}{\tau^2} - \frac{4 \xi k_*}{\tau}\sin(2 k_* \tau) + 4 \xi k_*^2 \cos(2 k_* \tau)+\mathcal{O}(\xi^2).
\end{equation}
Focusing on the sub-Hubble radius scenario, where $|k_* \tau| \gg 1$, the effective mass reduces to $z''/z = 4 \xi k_*^2 \cos(2 k_*\tau)$. The perturbation~\eqref{eq:uk} can then be rewritten as
\begin{equation}
    \frac{d^2 u_k}{dx^2} + \left( A_k - 2q \cos 2x \right) u_k =0,
\end{equation}
where we introduce the variable $x \equiv - k_* \tau$, $A_k = (1 - 2 \xi) k^2 / k_*^2$ and $q = (2 - k^2 / k_*^2) \xi$. This equation resembles the Mathieu equation, which exhibits parametric instability within specific ranges of $k$. Given that $\xi \ll 1$ and $|q| \ll 1$, the resonance bands are located in narrow ranges around $k \simeq n k_*$, where $n$ is an integer.

Let us focus on the first resonance band, corresponding to $n = 1$, as a representative example. At the onset of resonance, we initialize the value of $u_k$ to the Bunch-Davies vacuum, given by $u_k(\tau_0)=e^{-i k \tau_0}/ \sqrt{2k}$. For modes lying outside the resonance bands, $u_k$ remains nearly constant within the Hubble radius. In contrast, for modes within the resonance bands, the evolution of $u_k$ inside the Hubble radius can be characterized by the following expression:
\begin{equation}\label{eq:uk_tau}
    u_k(\tau) \propto \exp (\xi k_* \tau /2).
\end{equation}
This exponential behavior leads to a significant amplification of the perturbation modes within the resonance bands. Utilizing the relation $u \equiv z \zeta$ and \Eq{eq:uk_tau}, we can approximate the evolution of the comoving curvature perturbation for the $k_*$ mode as follows:
\begin{equation}
    \zeta_{k_*}(\tau) \simeq \zeta_{k_*}(\tau_0) e^{ \xi k_* (\tau - \tau_0)/2} \frac{\tau}{\tau_0}.
\end{equation}

This approximation is valid within the Hubble radius, where the initial value of the curvature perturbation, $\zeta_{k_*}(\tau_0) = -H \tau_0 /\sqrt{4 \epsilon k_*}$, is determined by the Bunch-Davies vacuum. As the mode crosses the Hubble radius at $\tau_*=-1/k_*$, the amplitude is significantly enhanced, reaching a value of
\begin{equation}
    \zeta_{k_*} \simeq \zeta_{k_*}(\tau_0) \left( \frac{-1}{k_* \tau_0} \right) e^{-\xi k_*\tau_0/2} \simeq \frac{H}{\sqrt{4 \epsilon k_*^3}} e^{-\xi k_* \tau_0/2}.
\end{equation}

The power spectrum of the primordial curvature perturbation, symbolized by $P_\zeta$, is defined as $P_\zeta \equiv k^3 |\zeta_k|^2 /(2 \pi^2)$. In the context of SSR, the resonance frequency undergoes an exponential amplification, leading to the formation of distinct peaks in the power spectrum at specific scales corresponding to the resonance bands. As a consequence of the exponential amplification, the complete power spectrum can be parametrized as follows:
\begin{equation}\label{Pk0}
    P_\zeta(k) = A_s \left( \frac{k}{k_p} \right)^{n_s-1} \left\{ 1+\frac{\xi k_*}{2}e^{-\xi k_*\tau_0} \left[ \delta(k-k_*) + \sum_{n=2}^{\infty} a_n \delta(k - n k_*) \right] \right\},
\end{equation}
where $A_s = H^2 / (8 \pi^2 \epsilon) \simeq 2.2\times 10^{-9}$ represents the amplitude of the power spectrum in standard inflation, $n_s \simeq 0.965$ corresponds to the spectral index at the pivot scale $k_p \simeq 0.05 ~ \text{Mpc}^{-1}$~\cite{Planck:2018vyg}, and $a_n \ll 1$ denotes the amplitude of the $n$th peak relative to the first peak. The coefficient preceding the $\delta$-function, $(\xi k_*/2)e^{-\xi k_*\tau_0}$, is determined by employing a triangle approximation to estimate the area of the peak. This approximation provides a simple and effective way to quantify the amplitude of the resonance peaks in the power spectrum, capturing the essential features of the amplified perturbations within the resonance bands. 

In this paper, we focus solely on the first peak, located at the characteristic scale $k_*$, while neglecting the contributions from higher-order peaks. This choice is justified by the fact that the amplitudes of the peaks at higher orders are exponentially suppressed, making their impact on the power spectrum significantly less pronounced compared to the first peak. By concentrating on the first peak, we can capture the dominant effects of sound speed resonance on the primordial curvature perturbations and simplify the analysis without sacrificing essential physical insights.

In the framework of the Newtonian gauge, the perturbed metric takes the following form:
\begin{equation}
ds^2 = a^2 \left\{-(1+2\Phi)\mathrm{d}\tau^2+[(1-2\Phi)\delta_{ij}+h_{ij}]\mathrm{d}x^i \mathrm{d}x^j\right\},
\end{equation}
where $\Phi$ represents the Bardeen potential, which characterizes the scalar perturbations, and $h_{ij}$ denotes the tensor perturbations. We neglect the contributions from first-order GWs, vector perturbations, and anisotropic stress, as they are considered subdominant~\cite{Baumann:2007zm,Weinberg:2003ur,Watanabe:2006qe}. 
{The Bardeen potential in the Fourier space, $\Phi_{\bm{k}}$, can be connected to the comoving curvature perturbations $\zeta_{\bm{k}}$ in the radiation-dominated era through the transfer function,
 \begin{equation}
 	\Phi_{\bm{k}}=\frac{2}{3}T(k\tau) \zeta_{\bm{k}},
 \end{equation}
where  the transfer function  $T(k\tau)$ satisfy
\begin{equation}\label{transfer}
T(z)=\frac{9}{z^2}\left[\frac{\sin (z / \sqrt{3})}{z / \sqrt{3}}-\cos (z / \sqrt{3})\right],
\end{equation}
with $z=k\tau$.}
Following the formalism presented in Refs.~\cite{Kohri:2018awv,Espinosa:2018eve}, the energy density of SIGWs at the epoch of matter-radiation equality can be expressed as
\begin{equation}
\Omega_{\mathrm{GW}}(k) = \int_0^{\infty} \mathrm{d} v \int_{|1-v|}^{1+v} \mathrm{d} u \mathcal{T}(u, v) {P}_{\zeta}(v k) {P}_{\zeta}(u k),
\end{equation}
where the {kernel} function $\mathcal{T}(u,v)$ is given by
\begin{equation}
\begin{aligned}
\mathcal{T}(u,v)= & \frac{3}{1024 v^8 u^8}\left[4 v^2-\left(v^2-u^2+1\right)^2\right]^2\left(v^2+u^2-3\right)^2 \\
& \times\bigg\{\left[\left(v^2+u^2-3\right) \ln \left(\left|\frac{3-(v+u)^2}{3-(v-u)^2}\right|\right)-4 v u\right]^2 \\
& +\pi^2\left(v^2+u^2-3\right)^2 \Theta(v+u-\sqrt{3})\bigg\}.
\end{aligned}
\end{equation}

{In the SSR model, the power spectrum can be approximated as a delta spectrum, $P_{\zeta}(k) \approx A \delta(\ln k- \ln k_*)$, where $A=0.5 A_s \xi e^{-\xi k_*\tau_0} \left(k_*/k_p \right)^{n_s-1}$. In such a case, the energy density of SIGWs, $\Omega_{\mathrm{GW}}(k)$, has an analytical form of 
\begin{equation}
\Omega_{\mathrm{GW}}(k)=\frac{A^2 \mathcal{T}\left(\frac{1}{\tilde{k}},\frac{1}{\tilde{k}}\right)}{\tilde{k}^2}\Theta(2-\tilde{k}),
\end{equation}
where $\tilde{k} \equiv k / k_*$ is the dimensionless wavenumber.}
By utilizing the relation between the wave number $k$ and frequency $f$, given by $k = 2\pi f$, the energy density fraction spectrum of SIGWs at the current epoch can be expressed as
\begin{equation}\label{omega:gw}
\Omega_{\mathrm{GW}, 0}(f)=\Omega_{\mathrm{r}, 0}\left[\frac{g_{, r}(T)}{g_{, r}\left(T_{\mathrm{eq}}\right)}\right]\left[\frac{g_{, s}\left(T_{\mathrm{eq}}\right)}{g_{, s}(T)}\right]^{\frac{4}{3}} \Omega_{\mathrm{GW}}(k),
\end{equation}
where $g_{,s}$ and $g_{,r}$ represent the effective degrees of freedom for entropy and radiation, respectively, and $\Omega_{r,0}$ is the present energy density fraction of radiation. 

\section{\label{data}Data analysis}
In this section, we outline the methodology used to constrain the SGWB with the SSR effect by analyzing the GW data from the first three observing runs of the Advanced LIGO and Virgo detectors. The LIGO-Hanford, LIGO-Livingston, and Virgo detectors form a network, with each detector labeled by the index $I={H, L, V}$. The time-series output of each detector, $s_I(t)$, is transformed into the frequency domain using a Fourier transform, yielding $\tilde{s}_I(f)$.

To search for the SGWB signal, we employ the cross-correlation statistic $\hat{C}^{I J}(f)$ for each detector pair (baseline) $IJ$, as~\cite{Romano:2016dpx,Allen:1997ad}
\begin{equation}\label{CIJ}
\hat{C}^{I J}(f)=\frac{2}{T} \frac{\operatorname{Re}[\tilde{s}_I^{\star}(f) \tilde{s}_J(f)]}{\gamma_{I J}(f) S_0(f)}.
\end{equation}
Here, $T$ is the observation time, $\gamma_{I J}(f)$ is the normalized overlap reduction function~\cite{Allen:1997ad} that accounts for the geometric sensitivity of the detector pair, and $S_0(f)=(3 H_0^2) /(10 \pi^2 f^3)$ is a normalization factor related to the critical energy density of the Universe. The cross-correlation statistic is normalized such that its expectation value equals the GW energy density spectrum, namely $\langle\hat{C}^{I J}(f)\rangle=\Omega_{\mathrm{GW}}(f)$, in the absence of correlated noise between the detectors. In the limit of a weak SGWB signal, the variance of the cross-correlation statistic can be approximated as
\begin{equation}
\sigma_{I J}^2(f) \approx \frac{1}{2 T \Delta f} \frac{P_I(f) P_J(f)}{\gamma_{I J}^2(f) S_0^2(f)},
\end{equation}
where $\Delta f$ is the frequency resolution, and $P_I(f)$ is the one-sided power spectral density of the noise in detector $I$. The variance $\sigma_{I J}^2$ allows us to estimate the uncertainty in the cross-correlation measurement based on the detector noise properties and the observation time.

\begin{table}[tbp]
\centering
\begin{tabular}{|c|c|c|}
\hline
\textbf{Parameter}& \textbf{Description} & \textbf{Prior} \\
\hline
$f_*\, (\mathrm{Hz})$ & Characteristic frequency & LogU$[10^{0}, 10^5]$ \\
$\xi$ & Amplitude of the oscillation & LogU$[10^{-8}, 0.25]$ \\
$|\tau_0|$ & Start time of the oscillation & LogU$[10^{-4}, 10^{9}]$ \\
\hline
\end{tabular}
\caption{\label{tab:prior}Prior distributions for the parameters of the SGWB model arising from the SSR effect. Here, LogU denotes the log-uniform distribution.}
\end{table}

We employ a Bayesian analysis to search for the SGWB signal originating from the SSR effect during inflation. The analysis utilizes the publicly available, model-independent cross-correlation spectra $\hat{C}^{I J}(f)$ data~\cite{KAGRA:2021kbb} from the first three observing runs of Advanced LIGO and Advanced Virgo detectors.
To estimate the parameters of the SGWB model arising from the SSR effect, we construct a likelihood function by combining the cross-correlation spectra from all detector pairs $I J$~\cite{Mandic:2012pj}
\begin{equation}\label{like}
p(\hat{C}^{I J}(f_k) | \boldsymbol{\theta}) \propto \exp \left[-\frac{1}{2} \sum_{I J} \sum_k\left(\frac{\hat{C}^{I J}(f_k)-\Omega_{\mathrm{M}}(f_k | \boldsymbol{\theta})}{\sigma_{I J}^2(f_k)}\right)^2\right],
\end{equation}
where $\boldsymbol{\theta}$ represents the set of parameters characterizing the SGWB model, denoted by $\Omega_{\mathrm{M}}(f | \boldsymbol{\theta})$. The likelihood assumes that the cross-correlation spectra $\hat{C}^{I J}(f_k)$ follow a Gaussian distribution in the absence of a signal. The sum runs over all frequency bins $k$ and detector pairs $I J$, with $\sigma_{I J}^2(f_k)$ being the variance of the cross-correlation statistic at each frequency bin.

Using Bayes' theorem, we express the posterior distribution of the model parameters as $p(\boldsymbol{\theta} | C_k^{I J}) \propto p(C_k^{I J} | \boldsymbol{\theta}) p(\boldsymbol{\theta})$, where $p(\boldsymbol{\theta})$ is the prior distribution on the parameters. To efficiently explore the parameter space and estimate the model parameters, we employ a Markov chain Monte Carlo (MCMC) method within the Bayesian framework. The MCMC algorithm generates samples from the posterior distribution, allowing us to infer the most likely parameter values and their uncertainties based on the observed data. This Bayesian approach provides a robust and flexible framework for constraining the parameters of the SGWB model arising from the SSR effect, taking into account the detector sensitivities and the statistical properties of the data.

\begin{figure}[tbp]
\centering
\includegraphics[width=0.9\textwidth]{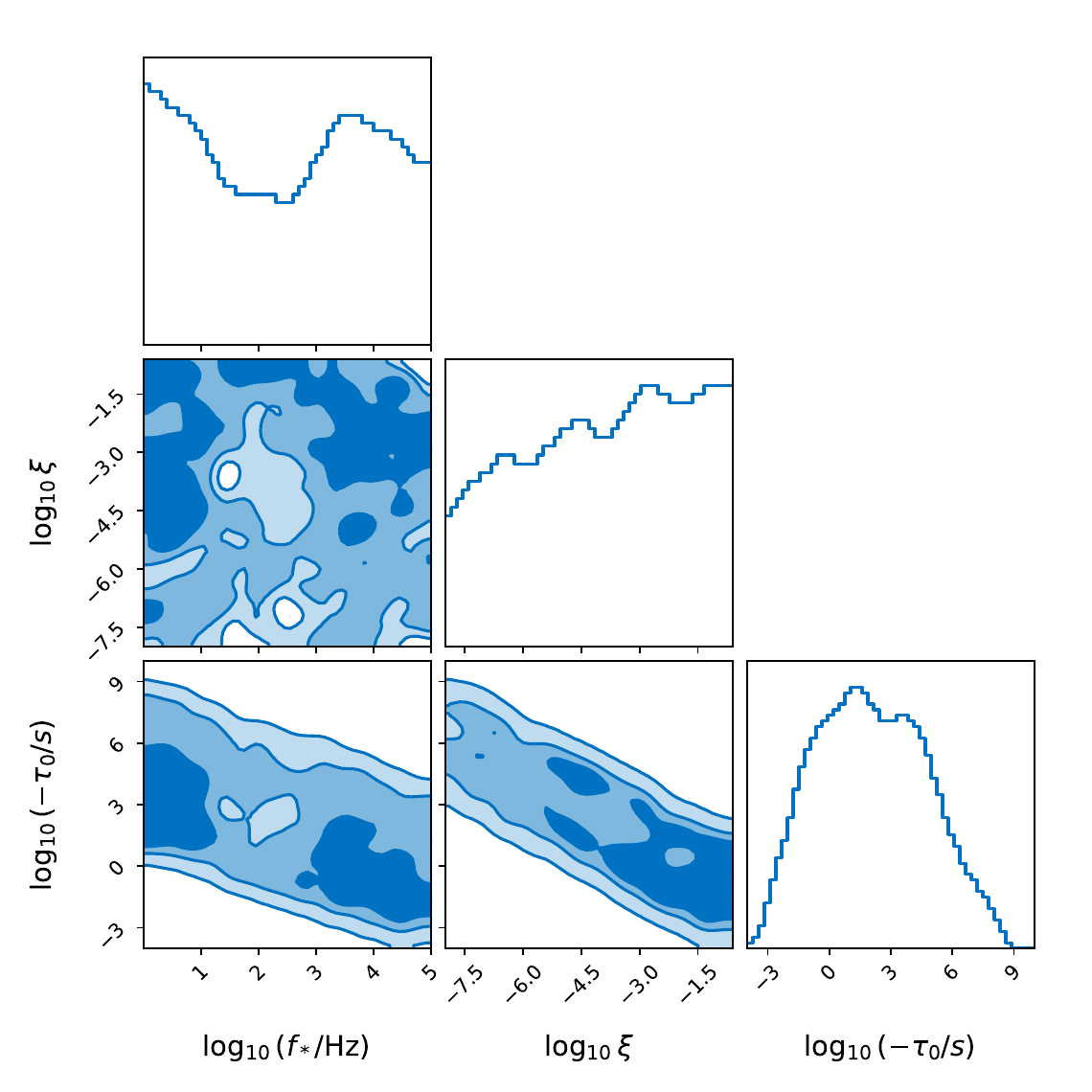}
\caption{\label{fig:posts}Posterior distributions for the parameters of the SGWB model from the SSR effect: characteristic frequency $f_*$, oscillation amplitude $\xi$, and start time of the oscillation $\tau_0$. The diagonal panels show the marginalized one-dimensional posteriors, while the off-diagonal panels display the joint two-dimensional posteriors with {confidence contours enclosing 68.27\% (1$\sigma$), 95.45\% (2$\sigma$), and 99.73\% (3$\sigma$) of the posterior probability}.}
\end{figure}

To assess the statistical significance of the presence of the SGWB signal from the SSR effect, we compute the Bayes factor, which compares the evidence for the model containing the SGWB signal to that of the model containing only noise
\begin{equation}
\mathcal{B}_{\mathrm{NOISE}}^{\mathrm{GW}}=\frac{p(\hat{C}^{IJ} | \text{Model with SGWB signal})}{p(\hat{C}^{IJ} | \text{Pure noise model})} = \frac{\int p(\hat{C}^{IJ}| \boldsymbol{\theta}_{\mathrm{GW}})\, p(\boldsymbol{\theta}_{\mathrm{GW}})\, \mathrm{d} \boldsymbol{\theta}_{\mathrm{GW}}}{\mathcal{N}}.
\end{equation}
The numerator represents the marginal likelihood of the model with the SGWB signal, obtained by integrating the product of the likelihood $p(\hat{C}^{IJ}| \boldsymbol{\theta}_{\mathrm{GW}})$ and the prior $p(\boldsymbol{\theta}_{\mathrm{GW}})$ over the parameter space $\boldsymbol{\theta}_{\mathrm{GW}}$. The denominator $\mathcal{N}$ is the evidence for the pure noise model, which is evaluated by setting $\Omega_{\mathrm{M}}(f)=0$ in \Eq{like}.
A Bayes factor $\mathcal{B}_{\mathrm{NOISE}}^{\mathrm{GW}}>1$ indicates support for the model with the SGWB signal compared to the pure noise model. The strength of the evidence can be interpreted using a standard scale, such as the Jeffreys scale~\cite{Jeffreys:1939xee}, where $\mathcal{B}_{\mathrm{NOISE}}^{\mathrm{GW}}>3, 10, 30, 100$ correspond to substantial, strong, very strong, and decisive evidence, respectively.

The free parameters in our analysis are $\boldsymbol{\theta}_{\mathrm{GW}} \equiv (f_*, \xi, \tau_0)$, where $f_*$ is the characteristic frequency, $\xi$ is the oscillation amplitude, and $\tau_0$ is the start time of the oscillation in the SSR model. The priors for these parameters are listed in \Table{tab:prior}.
We perform the Bayesian analysis using the \texttt{Bilby} package~\cite{Ashton:2018jfp,Romero-Shaw:2020owr}, which is a general-purpose Bayesian inference library. For sampling the parameter space, we employ the dynamic nested sampling algorithm implemented in the \texttt{Dynesty} package~\cite{Speagle:2019ivv}. Nested sampling is an efficient method for computing the marginal likelihood and estimating the posterior distributions of the parameters.

\begin{figure}[tbp]
\centering
\includegraphics[width=0.9\textwidth]{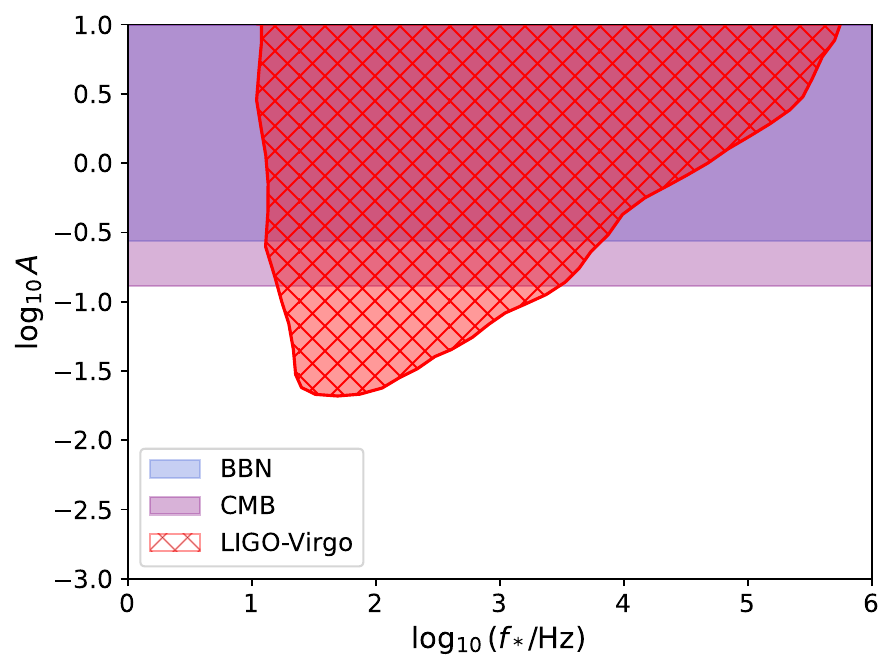}
\caption{\label{fig:cons}{Observational constraints on the primordial power spectrum in the amplitude-frequency plane. The shaded regions show excluded parameter space at 95\% confidence level from: LIGO-Virgo O1-O3 runs (red), big bang nucleosynthesis (BBN)~\cite{Cooke:2013cba} (blue), and cosmic microwave background (CMB)~\cite{Planck:2018vyg} (purple) observations.}}
\end{figure}

\section{\label{conclusion}Results and discussions}

In this work, we have conducted a search for the SGWB signal arising from SIGWs with the SSR effect. Our analysis utilized the combined cross-correlation spectrum from the first three observing runs of Advanced LIGO and Advanced Virgo, spanning a frequency range from 10 to 200 Hz. The SSR effect, characterized by an oscillating sound speed squared term, can lead to a nonperturbative parametric amplification of specific perturbation modes during inflation, resulting in enhanced primordial curvature perturbations and a significant SIGW signal.

We performed a Bayesian analysis to constrain the parameters of the SSR model, namely the characteristic frequency $f_*$, the oscillation amplitude $\xi$, and the start time of the oscillation $\tau_0$. The posterior distributions of these parameters are presented in Figure~\ref{fig:posts}. Our results indicate that the current instrumental capabilities of Advanced LIGO and Virgo are not sufficient to precisely determine the values of $f_*$ and $\xi$. This limitation arises from the fact that the expected SGWB signal from the SSR effect is likely to be weak and challenging to detect given the current detector sensitivities.

To quantify the statistical significance of the presence of the SGWB signal from the SSR effect, we computed the Bayes factor $\mathcal{B}_{\text{NOISE}}^{\text{GW}}$, which compares the evidence for the model with the SGWB signal to that of the pure noise model. Our analysis yielded a Bayes factor of $\mathcal{B}_{\text{NOISE}}^{\text{GW}}=0.83$, indicating that there is no significant evidence for the presence of such a signal in the current data. 

In the absence of a clear detection, we placed an upper limit on the start time of the oscillation, $\tau_0$, which is a key parameter in the SSR model. At a $95\%$ confidence level, we obtained a constraint of $|\tau_0| \lesssim 5.9\times 10^3$ seconds. This limit provides valuable information about the possible time scale of the SSR effect and can guide future theoretical and observational efforts in this direction. {Furthermore, we present the excluded regions of the primordial power spectrum in the amplitude-frequency plane in Fig.\ref{fig:cons}. The LIGO-Virgo O1-O3 observations provide stringent constraints at intermediate scales, excluding a significant portion of the parameter space at 95\% confidence level. These constraints are more stringent than those derived from BBN and CMB observations, as previously analyzed in Ref.\cite{Romero-Rodriguez:2021aws}.}

The SSR mechanism, which leads to resonating primordial curvature perturbations with an oscillatory feature in the sound speed of their propagation, can be realized within the framework of effective field theory of inflation or through non-canonical models inspired by string theory~\cite{Cai:2018tuh}. Our findings demonstrate the potential of using Advanced LIGO and Advanced Virgo data to constrain inflationary models that incorporate the SSR effect. This opens up new avenues for probing the early Universe and gaining insights into the physics of inflation.

In conclusion, while we did not find significant evidence for the presence of an SGWB signal from the SSR effect in the current data, our analysis sets the stage for future searches with improved detector sensitivities. As the advanced GW detectors continue to enhance their performance and new detectors come online, we anticipate that the constraints on the SSR model parameters will become more stringent. This work highlights the importance of searching for SGWB signals from novel mechanisms like the SSR effect, as they can provide unique opportunities to test theories of the early Universe and explore new physics beyond the standard inflationary paradigm.

\section*{Acknowledgments}
YW is supported by the National Natural Science Foundation of China under Grant No.~12405057.
ZCC is supported by the National Natural Science Foundation of China under Grant No.~12405056 and the innovative research group of Hunan Province under Grant No.~2024JJ1006.
LL is supported by the National Natural Science Foundation of China Grant under Grant No. 12433001.  

\bibliographystyle{JHEP}
\bibliography{ref}
\end{document}